\documentclass{article}

\usepackage{spconf,amsmath,graphicx,xcolor,acronym, comment, color,cite}
\usepackage{booktabs}
\usepackage{numprint}
\usepackage{colortbl}
\usepackage{hyperref}
\usepackage{multirow}
\usepackage{multicol}

\title{Adversarial speech for voice privacy protection from Personalized Speech generation}

\name{Shihao Chen$^1$, Liping Chen$^{1,*}$, Jie Zhang$^1$, KongAik Lee$^2$, Zhenhua Ling$^1$, Lirong Dai$^1$\thanks{$^*$Corresponding author}}

\address{$^1$NERC-SLIP, University of Science and Technology of China (USTC), China\\$^2$Department of Electrical and Electronic Engineering, The Hong Kong Polytechnic University, Hong Kong \\
\small \tt shchen16@mail.ustc.edu.cn lipchen@ustc.edu.cn}

\begin{document}
\ninept
\maketitle
\begin{abstract}
The rapid progress in personalized speech generation technology, including personalized text-to-speech (TTS) and voice conversion (VC), poses a challenge in distinguishing between generated and real speech for human listeners, resulting in an urgent demand in protecting speakers' voices from malicious misuse. In this regard, we propose a speaker protection method based on adversarial attacks. The proposed method perturbs speech signals by minimally altering the original speech while rendering downstream speech generation models unable to accurately generate the voice of the target speaker. For validation, we employ the open-source pre-trained YourTTS model for speech generation and protect the target speaker's speech in the white-box scenario. Automatic speaker verification (ASV) evaluations were carried out on the generated speech as the assessment of the voice protection capability. Our experimental results show that we successfully perturbed the speaker encoder of the YourTTS model using the gradient-based I-FGSM adversarial perturbation method. Furthermore, the adversarial perturbation is effective in preventing the YourTTS model from generating the speech of the target speaker. Audio samples can be found in \url{https://voiceprivacy.github.io/Adeversarial-Speech-with-YourTTS}. 
\end{abstract}
\begin{keywords}
Personalized speech generation, text-to-speech, voice conversion, voice privacy, adversary attack.
\end{keywords}
\section{Introduction}
\label{sec:intro}
In the era of deep learning, techniques in speech generation have developed rapidly, greatly improving the quality and naturalness of the artificial speech. Among them, the personalized speech generation techniques, including \emph{text-to-speech synthesis} (TTS) \cite{casanova2022YourTTS} and \emph{voice conversion} (VC) \cite{9262021} facilitate generating the speech of a target speaker in high speaker similarity, raising the potential security risks concerning the voice privacy. To be specific, given the speech utterances of a target speaker for reference, multi-speaker TTS and VC technologies can be utilized to generate its speech utterances with fabricated content, which can potentially be used to spoof a speaker authentication system or manipulate public opinions.

How to protect the voice privacy in the speech utterances from being utilized to generate speech for malicious purposes has attracted the interest of the research community. In this field, several approaches have been explored. To name a few, driven by the ASVspoof challenges \cite{liu2023asvspoof}, the synthesized speech detection techniques have developed remarkably, relieving the threats of synthesized speech on speaker authentication systems. In addition, the concept of anonymization provides another voice privacy protection method by concealing the speaker attributes in the speech utterances before they are used for malicious speech generation. In this manner, the attributes of the original speaker will not be utilized by speech modeling techniques. Moreover, it also hides the original speaker attributes from human listening. Between these two approaches, the former works on passive protection where the detection is carried out after the artificial speech of the target speaker is generated; the latter operates proactively by preventing the artificial speech generation.

This paper follows the route of proactively protecting voice privacy and aims at preventing speech from being utilized in personalized speech generation. In contrast to existing anonymization methods that modify speaker attributes and alter human perception, our study pursues that the voices in both the original and anonymized speech utterances sound alike to human listeners. To achieve this goal, the adversarial attack concept will be applied and the adversarial speech will be generated, within which the voice attributes are protected. Adversarial perturbation on neural networks was discovered in \cite{szegedy2014intriguing} where the neural network models were found to be misled by adversarial samples. Based on the findings, adversarial perturbation has been exploited to attack the models of face image recognition \cite{aneja2022tafim}, speech recognition\cite{neekhara2019universal}, speaker identification\cite{yao2023symmetric,shamsabadi2021foolhd}, speaker verification \cite{zhang2021attack,kreuk2018fooling} and so on, successfully misleading the models to make wrong predictions without altering the perceptions of the original face images and speech utterances.

Currently, in personalized speech generation models, the speaker modeling is mainly built on extracting speaker attributes with a separate speaker encoder, e.g., \cite{casanova2022YourTTS,li2023freevc}. Given the speech utterances of a target speaker, the adversarial attack will be carried out on the speaker encoder to obtain the adversarial speech utterances. In this manner, the speaker encoder will become incapable of extracting the speaker attributes of the target speaker from the adversarial utterances. As a result, the speech generation model cannot generate the speech of the target speaker using the adversarial speech, finally, protecting the voice privacy. As a preliminary work, we will focus on white-box scenario, i.e., preventing a known speech generation model from generating the speech of a target speaker. YourTTS \cite{casanova2022YourTTS} is selected to be our white-box speech generation model, which can perform both zero-shot TTS and VC. Given a well-trained YourTTS model, the adversarial attack methods based on \emph{fast gradient step method} (FGSM) \cite{goodfellow2014explaining} are applied to attack the speaker encoder and thus generate the adversarial speech. To this end, a loss function is defined on the speaker encoder as the cosine similarity between the speaker embeddings extracted from the original utterance and its adversarial sample. Evaluations are carried out on the generated speech of the target speaker using the adversarial speech as a reference for speaker attributes extraction. The \emph{equal error rate} (EER) obtained in automatic speaker verification (ASV) is adopted as the performance metric, measuring the effectiveness of the adversarial speech utterance in protecting the true speaker attributes from being perceived by the speaker encoder.

The rest of this paper is organized as follows. In Section \ref{sec:background}, we briefly introduce the background of our work, including the YourTTS model and the FGSM methods. Section \ref{sec:adversarial speech} presents the proposed method for adversarial speech generation. The experiments will be presented in Section \ref{sec:experiments}. Finally, Section \ref{sec:conclusion} concludes this work.

\section{background}
\label{sec:background}

\begin{figure}[!]
  \centering
  \includegraphics[width=3in]{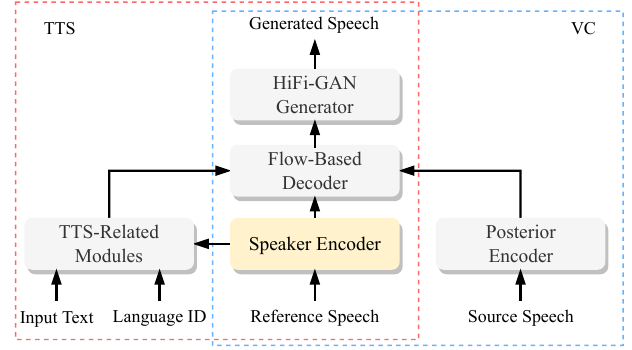}
  \caption{The inference flowchart of the YourTTS model. The blue and red boxes represent the flows of TTS and VC, respectively.}
  \label{yourtts}
\end{figure}

In this section, we will first briefly introduce the inference process of YourTTS model for personalized speech generation. Then we will describe the adversarial attack approaches that will be used in our work, including the FGSM and \emph{iterative FGSM} (I-FGSM).
\subsection{YourTTS}
Fig. \ref{yourtts} illustrates the inference process of YourTTS. As shown in Fig. \ref{yourtts}, YourTTS is an end-to-end model that can be applied to generate personalized speech in both TTS and VC manners. The TTS flow takes the raw text and the language ID as input. The inputs are processed by a sequence of TTS-related modules. To be specific, as detailed in \cite{casanova2022YourTTS}, the TTS-related module is composed of an encoder followed by an alignment generation module. On the VC side, the model takes the source speech as input and processes it with a posterior encoder. Finally, a flow-based decoder followed by a HiFi-GAN \cite{kong2020hifi} generator is applied to generate the speech in both the TTS and VC flows. Additionally, in both the TTS and VC flows, a reference speech spoken by the target speaker is applied to provide the information of the target speaker. The speaker attributes are extracted with the speaker encoder and utilized in the TTS-related modules and the HiFi-GAN module to generate the speech of the target speaker. For details, readers are suggested to read \cite{casanova2022YourTTS}.

 In our work, the open-source YourTTS model\footnote{\url{https://github.com/Edresson/YourTTS}} will be used whose speaker encoder utilizes the H/ASP \cite{heo2020clova} structure, which is an improved version of ResNet-34 \cite{zeinali2019but}. The H/ASP model was trained on the VoxCeleb2 dataset \cite{chung2018voxceleb2} and obtained an EER of 1.967\% on the test subset of the multi-language LibriSpeech (MLS) dataset \cite{pratap2020mls}.

\subsection{FGSM \& I-FGSM}
FGSM \cite{goodfellow2014explaining} is an adversarial attack method based on the optimization of neural networks. Denote the loss function as $L\left(f\left({\bf x}\right),{\bf y}^{\rm true}\right)$, where $f\left(\bullet\right)$ represents the network function, ${\bf x}$ and ${\bf y}^{\rm true}$ are the original sample and its ground truth target, respectively. The adversarial sample of ${\bf x}$ is obtained with:
\begin{equation}
\label{eq. FGSM}
    \widetilde{{\bf x}} = {\bf x} + \epsilon\cdot sign\left(\nabla_{\bf x} L\left(f\left({\bf x}\right),{\bf y}^{\rm true}\right)\right),
\end{equation}
where $\nabla_{\bf x}L\left(\bullet\right)$ is the derivative of the loss function $L$ on $\bf x$. Besides, $sign\left(\bullet\right)$ is the sign operation, $\epsilon$ is the attacking intensity parameter, and ${\widetilde{\bf x}}$ is the adversarial sample of ${\bf x}$.

FGSM is a single-step optimation approach. Furthermore, I-FGSM provides an improved version by incorporating the iterative training mechanism. In I-FGSM \cite{kurakin2018adversarial}
, the adversarial sample ${\bf \widetilde{x}}$ is generated with:
\begin{equation}
\label{eq. BIM 2}
    {{\bf{\tilde x}}_{i + 1}} = {\rm{Clip}}^\epsilon \left\{{{\bf{\tilde x}}_i} + \underbrace{\alpha  \cdot sign \left( {{\nabla _{{\bf{\tilde x}}_i}} \left( { L\left( f\left( {\bf{\tilde x}}_i \right)   ,{{\bf y}^{\rm{true}}} \right)}  \right)} \right) }_{\delta_i} \right\} ,
\end{equation}
where $i=0,1,2...,I-1$ indicates the iterations with $I$ to be the number of iterations. In (\ref{eq. BIM 2}), ${{{\bf{\tilde x}}}_0}$ is initialized to ${\bf{x}}$, $0<\alpha<\epsilon$ is the step size smaller than the $\epsilon$ in (\ref{eq. FGSM}), and $\delta_i$ denotes the adversarial perturbation as added to ${\bf {\tilde x}}_i$. The ${\rm Clip}^{\epsilon}\left(\bullet\right)$ function limits $\tilde{\bf x}$ the vicinity of $\bf x$ satisfying the $L_{\infty}$ norm bound.

\begin{figure}[!t]
  \centering
  \includegraphics[width=3in]{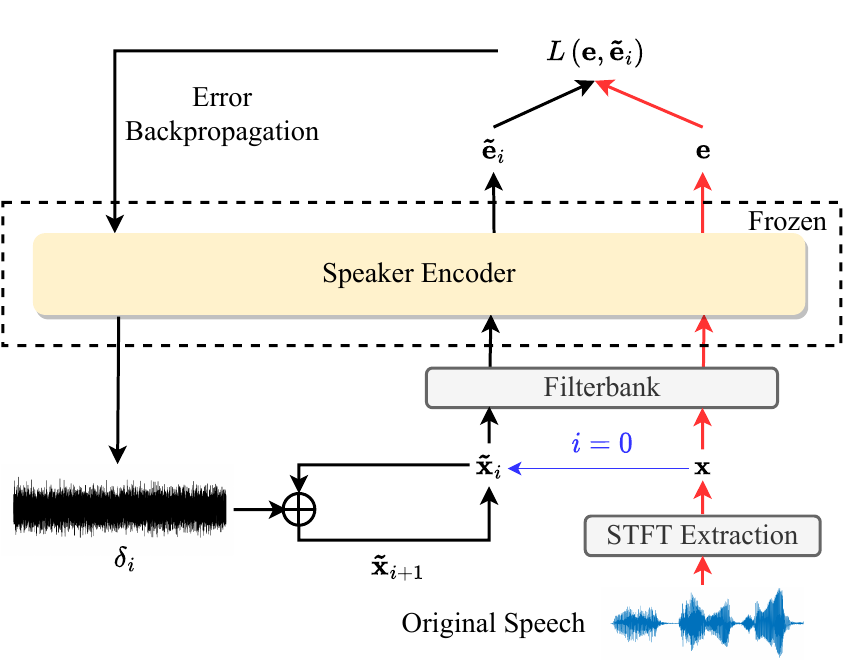}
  \caption{The flowchart of generating adversarial samples based on I-FGSM. The ${\bf x}$ denotes the STFT feature extracted from the original reference speech; $\delta_i$ and ${\bf \tilde{x}}_i$ represent the adversarial perturbation and the resultant adversarial STFT in the $i$-th iteration, respectively. When $i=0$, ${\bf \tilde{x}}_0$ is initialized with $\bf x$.}.
  \label{loss}
\end{figure}

\section{Speaker adversarial speech}
\label{sec:adversarial speech}
In our work, FGSM and I-FGSM are applied to generate the adversarial speech. Given a well-trained YourTTS model, Fig. \ref{loss} illustrates the adversarial speech generation process based on its speaker encoder using the I-FGSM method. The frequency bins as extracted with \emph{short-time Fourier transform} (STFT) are used for adversarial speech generation.

Assume the STFT extracted from the original speech to be ${\bf x}$ and I-FGSM is implemented for $I$ iterations. Firstly, the adversarial STFT ${\tilde{\bf x}}_0$ is initialized with ${\bf x}$. In the $i$-th iteration, the filterbank feature vectors are first extracted from ${\bf x}$ and ${\bf {\tilde x}}_i$, based on which the speaker embedding vectors are extracted with the speaker encoder, denoted as ${\bf e}$ and ${\bf {\tilde{e}}}_i$, respectively. As FGSM methods are based on error backpropagation, a loss function is exerted on the speaker embedding vectors. Mathematically, it is defined as the cosine distance between ${\bf e}$ and ${\bf \tilde{e}}_i$ as follows:

\begin{equation}
\label{eq. loss function}
    L\left({\bf{e}}, {\bf{\tilde e}}_i\right) =  -\frac{{\bf{e}}^{\mathsf T} {\bf{\tilde e}}_i}{\|{\bf{e}} \|_2 
  \|{\bf{\tilde e}}_i \|_2}.
\end{equation}
A single FGSM step is implemented via loss backpropagation to ${\tilde{\bf x}}_i$ to increase the cosine distance between ${\bf e}$ and ${\bf \tilde{e}}_i$. When the perturbation with respect to ${\bf {\tilde x}}_i$ is obtained as $\delta_i$, the adversarial STFT is updated by adding $\delta_i$ to ${\bf {\tilde x}}_i$ as follows:
\begin{equation}
    \label{eq. adversarial update}
    {{\bf {\tilde x}}_{i+1}} \leftarrow{{\bf \tilde x}_i+\delta_i}.
\end{equation}
After $I$ iterations, the adversarial STFT can be generated as ${\bf {\tilde x}}_{I}$. When $I=1$, the I-FGSM degrades to FGSM.

Upon ${\bf {\tilde x}}_{I}$, the \emph{inverse STFT} (iSTFT) is used to generate the adversarial speech using the phase as extracted from the original speech. Since the I-FGSM is carried out on the speaker encoder of the YourTTS model, the adversarial speech is expected to prevent it from extracting the correct speaker attributes. When used as reference speech, it makes the YourTTS model unable to generate the speech of the target speaker.

\section{Experiments}
\label{sec:experiments}

Our experiments were conducted on the test-clean subset of the Librispeech \cite{panayotov2015librispeech} dataset, denoted as \emph{libri-test} in the following. All the recordings were resampled to 16kHz. For adversarial speech generation, $512$ frequency bins were extracted with Hanning window, with a $10$ ms frameshift and a $25$ ms window size.

In our white-box protection experiments, the YourTTS model trained with the configuration of Exp.4 as illustrated in \cite{casanova2022YourTTS} was employed for both adversarial speech generation and evaluation. Specifically, the utilized YourTTS model was firstly trained using the multi-lingual corpus composed of VCTK \cite{yamagishi2019vctk} and their internal datasets, including TTS-Portuguese \cite{casanova2022tts}, and M-AILABS French (trilingual) \cite{malabs}. Then, the model was finetuned on the train-clean-100 and train-clean-360 datasets of LibriTTS \cite{zen2019libritts}. In our experiments, three types of perturbed speech were evaluated, including the Gaussian white noise and the adversarial speech generated with our proposed method where both FGSM and I-FGSM are applied. For FGSM, the attack intensity $\epsilon$ was set to $0.02$. In the experiments of I-FGSM, the number of iterations $I$ was set to $50$ with $\alpha=0.0004$.

The effectiveness of the adversarial speech was examined in both the zero-shot TTS and VC. To evaluate the speaker similarity among the generated speech utterances, the \emph{automatic speaker verification} (ASV) evaluations were carried out using the \emph{equal error rate} (EER) as the performance metric. In our ASV evaluations, the official trials on the libri-test dataset as provided by VoicePrivacy Challenge 2022 \cite{chen2022system} were adopted. The ASV evaluations were conducted on the speech generated by the YourTTS model. In TTS speech evaluation, given a pair of a recording and its transcript, the speech was synthesized with the transcript to be the input text and the recording as the reference. In the experiments on VC, the recording was used as the reference while the source speech was randomly selected from a different speaker. Similarly, in the experiments of perturbed speech, the perturbed utterances were used as references in YourTTS. The higher EER infers better voice privacy capability of the perturbation.

ECAPA-TDNN \cite{ECAPATDNN} combined with cosine distance was applied in our ASV evaluation. The open-source toolkit ASVsubtools toolkit\footnote{\url{https://github.com/Snowdar/asv-subtools}} was used to train the ECAPA-TDNN model on the Librispeech-360-clean dataset. In our experiments, we took into account the domain mismatch between the original recording and the generated speech. For evaluation purposes, we utilized two ASV models: one trained on the original recording, denoted as $ASV_{rec}$, and another trained on the TTS speech of Librispeech-360-clean, generated with the TTS function of the YourTTS model, denoted as $ASV_{tts}$. During the training of the two models, data augmentation using the MUSAN corpus \cite{snyder2015musan} and the RIR dataset \cite{ko2017study} was applied.

Firstly, the speech quality of the perturbed speech was evaluated on the libri-test evaluation dataset, with the results as presented in Table \ref{quality}. The metrics of \emph{perceptual evaluation of speech quality} (PESQ) \cite{hu2007evaluation} and \emph{signal-to-noise ratio} (SNR) were used to assess the distortion level of the perturbed samples. The results are presented in Table \ref{quality}. Besides, the cosine distance between the speaker embedding vectors as extracted with the speaker encoder in YourTTS model from the recording and the adversarial speech was computed according to (\ref{eq. loss function}). The averaged cosine distance across the utterances in libri-test is shown in Table \ref{quality}, denoted as $\Delta CosD$, which ranges from $-1$ to $1$. A higher value of $\Delta CosD$ indicates a lower level of similarity. It can be observed from Table \ref{quality} that I-FGSM achieved the highest SNR and PESQ, inferring its superiority to the Gaussian white noise and FGSM in preserving the perceptual quality of the recording. In the $\Delta CosD$ metric, I-FGSM also achieved the highest cosine distance between the speaker embedding vectors extracted from the recording and the perturbed speech, inferring the best speaker protection capability.
\begin{table}[!h]
\renewcommand{\arraystretch}{0.88}
    \centering   
    \caption{The PESQ, SNR and $\Delta CosD$ results on the perturbed speech obtained with Gaussian white noise, FGSM and I-FGSM.}
    \begin{tabular}{l|c|c|c}
        \toprule
        Method & SNR(dB)$(\uparrow)$ & PESQ$(\uparrow)$ & $\Delta CosD(\uparrow)$ \\ \midrule
        Gaussian & 32.02& 3.79  & -0.96\\ 
        FGSM\cite{goodfellow2014explaining} &  32.04 & 3.78 & -0.80\\
        I-FGSM\cite{kurakin2018adversarial}  & 37.00 & 4.17 & 0.06 \\
        \bottomrule
    \end{tabular}
    \label{quality}
\end{table}

\begin{table*}[]
\caption{The speaker verification EER (\%) on recording ($rec$) , voice conversion data ($vc$), text-to-speech data ($tts$), adversarial voice conversion data ($vc^*$) and adversarial text-to-speech data ($tts^*$). The performances are presented in the \emph{enrollment-trial} format. Four kinds of speech utterances used for reference in personalized speech generation are included: recording, recording with Gaussian white noise (Gaussian), adversarial speech obtained by FGSM (FGSM) and adversarial speech obtained by I-FGSM (I-FGSM).}
\renewcommand{\arraystretch}{0.88}
\setlength\tabcolsep{3pt} 
\centering
\begin{tabular}{@{}l|cccccccccccc@{}}
\toprule
\multirow{3}{*}{Method} & \multicolumn{6}{c|}{$ASV_{rec}$}                          & \multicolumn{6}{c}{$ASV_{tts}$}      \\ \cmidrule(l){2-13} 
                        & \multicolumn{12}{c}{Male}                                                         \\ \cmidrule(l){2-13} 
 &
  \multicolumn{1}{c}{$rec\mbox{-}tts^*$} &
  \multicolumn{1}{c}{$tts\mbox{-}tts^*$} &
  \multicolumn{1}{c|}{$tts^*\mbox{-}tts^*$} &
  \multicolumn{1}{c}{$rec\mbox{-}vc^*$} &
  \multicolumn{1}{c}{$vc\mbox{-}vc^*$} &
  \multicolumn{1}{c|}{$vc^*\mbox{-}vc^*$} &
  \multicolumn{1}{c}{$rec\mbox{-}tts^*$} &
  \multicolumn{1}{c}{$tts\mbox{-}tts^*$} &
  \multicolumn{1}{c|}{$tts^*\mbox{-}tts^*$} &
  \multicolumn{1}{c}{$rec\mbox{-}vc^*$} &
  \multicolumn{1}{c}{$vc\mbox{-}vc^*$} &
  $vc^*\mbox{-}vc^*$ \\ \midrule
recording              & 5.79 & 3.79 & \multicolumn{1}{c|}{3.79} & 10.24  & 8.46 & \multicolumn{1}{c|}{8.46} & 5.12 & 2.45 & \multicolumn{1}{c|}{2.45} & 8.69 & 5.35 & 5.35 \\
Gaussian                & 9.13 & 5.12 & \multicolumn{1}{c|}{3.79} & 12.03   & 9.80 & \multicolumn{1}{c|}{9.13} & 9.35 & 4.90 & \multicolumn{1}{c|}{2.90} & 11.80 & 8.24 & 6.46 \\
FGSM                    & 10.91 & 6.01 & \multicolumn{1}{c|}{5.79} & 15.14 & 12.69 & \multicolumn{1}{c|}{13.81} & 10.91 & 5.35 & \multicolumn{1}{c|}{4.90} & 14.92 & 9.80 & 10.24 \\
I-FGSM                  & \bf{48.11} & \bf{45.21} & \multicolumn{1}{c|}{\bf{21.83}} & \bf{49.89} & \bf{48.55} & \multicolumn{1}{c|}{\bf{28.95}} & \bf{52.12} & \bf{48.11} & \multicolumn{1}{c|}{\bf{16.04}} & \bf{52.12} & \bf{51.22} & \bf{24.94} \\ \midrule
                        & \multicolumn{12}{c}{Female}                                                           \\ \midrule
recording              & 9.31 & 7.48 & \multicolumn{1}{c|}{7.48} & 10.04 & 10.58 & \multicolumn{1}{c|}{10.58} & 8.03 & 7.48 & \multicolumn{1}{c|}{7.48} & 10.22 & 9.12 & 9.12\\
Gaussian                & 8.57 & 7.64 & \multicolumn{1}{c|}{6.02} & 11.68 & 11.13 & \multicolumn{1}{c|}{10.40} & 8.21 & 4.93 & \multicolumn{1}{c|}{4.92} & 11.68 & 8.76 & 7.30 \\
FGSM                    & 10.58 & 9.67 & \multicolumn{1}{c|}{9.12} & 13.50 & 13.69 & \multicolumn{1}{c|}{13.87} & 11.86 & 7.48 & \multicolumn{1}{c|}{4.93} & 14.60 & 11.31 & 11.86 \\
I-FGSM                  & \bf{51.28} & \bf{47.26} & \multicolumn{1}{c|}{\bf{23.91}} & \bf{50.73} & \bf{49.27} & \multicolumn{1}{c|}{\bf{28.65}} & \bf{52.55} & \bf{53.01} & \multicolumn{1}{c|}{\bf{21.53}} & \bf{52.37} & \bf{51.64} & \bf{28.28} \\ \bottomrule
\end{tabular}
\label{result_eer}
\end{table*}

Next, in order to evaluate the voice protection capability, ASV evaluations were carried out on the generated speech, where both TTS and VC were applied for speech generation. Table \ref{result_eer} shows the EERs obtained on the utterances as generated using YourTTS with the reference speech being the recording and perturbed speech utterances, respectively. In the evaluations, both the models trained on recording ($ASV_{rec}$) and TTS speech ($ASV_{tts}$) were used. In Table \ref{result_eer}, the terms \emph{tts} and \emph{vc} refer to the speech generated when the original recording is used as a reference in the TTS and VC speech generation processes, respectively. On the other hand, $tts^{*}$ and $vc^{*}$ indicate the speech generated when the perturbed speech was utilized as a reference. The Gaussian white noise, adversarial perturbation generated with FGSM and I-FGSM were examined. For comparison, the original recording was also experimented. Regarding the enrollment utterance in the evaluation trials, the recording (\emph{rec}), the speech generated with recording as reference ($tts$ and $vc$) and the speech generated with perturbed speech as reference ($tts^*$ and $vc^*$) were applied. $tts^*$ and $vc^*$ were used as the test utterances.

From the results as presented in Table \ref{result_eer}, we can see that despite the domain difference, $ASV_{rec}$ and $ASV_{tts}$ provided consistent EERs among the recording and the three types of perturbed utterances. It is noteworthy that the generated speech utterances $tts$ and $tts^*$ given recording as reference speech are identical. Both infer that the speech was generated by YourTTS with the recording being used as reference speech. This is also applicable to the $vc$ and $vc^*$ cases. Compared with the recording, when the perturbed utterances as obtained by the three methods were used for reference in YourTTS speech generation, higher EERs can be obtained in both TTS and VC flows in all trial conditions. This suggests that the perturbation techniques effectively perturbed the speaker attributes extraction by the speaker encoder, thus impairing the ability of the YourTTS model to generate the voice of the reference speaker. The adversarial speech generated with I-FGSM achieved the highest EERs in all trial conditions. Specifically, the $rec\mbox{-}tts^*$ trials resulted in approximately 50\% EERs, indicating that using the adversarial speech for reference is unable to generate the speech of the target speaker. Likewise, the nearly 50\% EERs observed in the $tts\mbox{-}tts^*$ trials suggest that the voices are different between the utterances as generated using the recording and adversarial speech for the reference of the target speaker. The same findings can be seen from the VC evaluations, i.e., the $rec\mbox{-}vc^*$ and $vc\mbox{-}vc^*$ scenarios. With respect to the $tts^*\mbox{-}tts^*$ and $vc^*\mbox{-}vc^*$ evaluations, EERs around 20\% were obtained by I-FGSM, which are lower than 50\%. This infers that since the adversarial utterances were generated without considering the speaker distinction within them, the speaker attributes as perceived by the model still present speaker clustering characteristics.

\begin{figure}[h]
  \centering
  \includegraphics[width=3in]{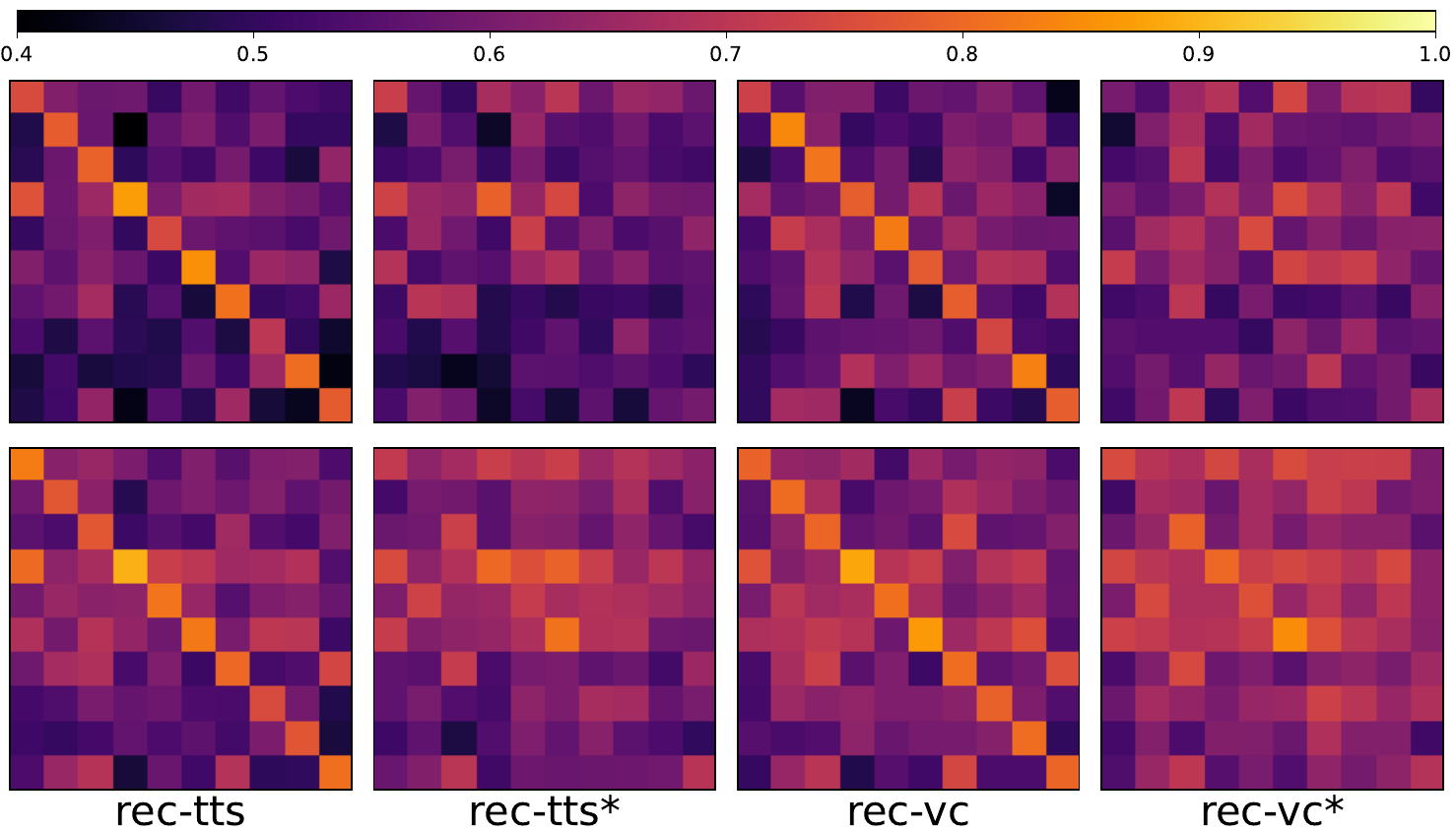}
  \caption{The similarity among utterances (first row) and speakers (second row). The speaker similarity between the recording ($rec$) and the speech generated using the recording as the reference speech ($tts$ and $vc$), using adversarial speech for reference ($tts^*$ and $vc^*$) are presented, respectively.}
  \label{visual}
\end{figure}

Finally, given the speech generated using recording and adversarial speech as the reference speech, respectively, the speaker similarity with the recording is visualized as given in Fig. \ref{visual}. In the visualization, the open-source toolkit Resemblyzer \cite{wan2018generalized} was applied where the cosine similarity between the speaker embedding vectors were computed as the similarity score. Five male and five female speakers were randomly selected from the libri-test dataset, with five randomly selected utterances for each speaker. The speaker similarity was visualized in utterances and speakers, respectively. The first and second rows of Fig.~\ref{visual} show the similarity among utterances and speakers, respectively. In computing the speaker similarity among the speakers, the embedding vector for each speaker was obtained by averaging that of the five utterances. In Fig \ref{visual}, the speaker similarity between the recording and the generated utterances including $tts$, $tts^*$, $vc$ and $vc^*$ are presented, respectively. It can be observed that $tts$ and $vc$ exhibit a relatively high similarity along the diagonal with respect to $rec$. This is consistent with the EERs observed in the $rec\mbox{-}tts^*$ and $rec\mbox{-}vc^*$ conditions when recording is utilized as reference speech in Table \ref{result_eer}. When the I-FGSM method is applied to protect the voice attributes, the speaker identity in the generated speech becomes unrelated to the speaker in the recording. This reveals the efficacy of the proposed method in protecting the speaker privacy when being used for personalized speech generation.

\section{Conclusion}
\label{sec:conclusion}

In this study, we introduced a method to protect speaker privacy through adversarial speech generation, aiming at preventing the use of speaker attributes for generating speech imitating a specific target speaker. We investigated adversarial sample generation methods based on FGSM in a white-box protection scenario. The open-source pre-trained YourTTS model was used for personalized speech generation, whose speaker encoder was attacked to generate adversarial speech utterances. Our experiments, conducted on the test-clean dataset of LibriSpeech, demonstrated the effectiveness of our approach in preventing the YourTTS model from imitating a target speaker when provided with adversarial speech as a reference.

\vfill\pagebreak

\bibliographystyle{IEEEbib}
\bibliography{strings,refs}

\end{document}